\documentclass[a4paper]{article}
\pdfoutput=1 
\usepackage{fullpage,epsfig,amsmath,amsfonts,amssymb,array,graphicx,wrapfig,float,caption,subcaption,xfrac,longtable,fancyhdr,textgreek}
\usepackage[doublespacing]{setspace}  
\usepackage{times}
\usepackage{pdfpages}

\usepackage[a4paper,top=3cm,bottom=2cm,left=3cm,right=3cm,marginparwidth=1.75cm]{geometry}
\usepackage{abstract}

\title{Imaging Through Noise With Quantum Illumination}

\author
{T. Gregory, P.-A. Moreau, E. Toninelli, and M.J. Padgett$^{\ast}$\\
	\normalsize{School of Physics and Astronomy, University of Glasgow,}\\
	\normalsize{Glasgow, G12 8QQ, UK}\\
	\normalsize{$^\ast$To whom correspondence should be addressed; E-mail: miles.padgett@glasgow.ac.uk.}
}
\date{}
\begin{document} 
	\maketitle 
	
\begin{abstract}
\noindent \textbf{The contrast of an image can be degraded by the presence of background light and sensor noise. To overcome this degradation, quantum illumination protocols have been theorized that exploit the spatial correlations between photon-pairs. Here, we demonstrate the first full-field imaging system using quantum illumination, by an enhanced detection protocol. With our current technology, we achieve a rejection of background and stray light of up to 5.8 and also report an image contrast improvement up to a factor of 11, which is resilient to both environmental noise and transmission losses. The quantum illumination protocol differs from usual quantum schemes in that the advantage is maintained even in the presence of noise and loss. Our approach may enable laboratory-based quantum imaging to be applied to real-world applications where the suppression of background light and noise is important, such as imaging under low -photon flux and quantum LIDAR.}                                                               \end{abstract}

\section*{Introduction}
Conventional illumination uses a spatially and temporally random sequence of photons to illuminate an object, whereas quantum illumination can use spatial correlations between pairs of photons to achieve performance enhancements in the presence of noise and/or losses. This enhancement is made possible by using detection techniques that preferentially select photon-pair events over isolated background events. The quantum illumination (QI) protocol was introduced by Lloyd \cite{lloyd_enhanced_2008}, and generalized to Gaussian states by Tan et al. \cite{tan_quantum_2008}, where they proposed a practical version of the protocol. Quantum illumination has applications in the context of quantum information protocol such as secure communication \cite{shapiro_defeating_2009,zhang_entanglements_2013} where it secures communication against passive eavesdropping techniques that take advantage of noise and losses. The protocol has also been proposed to be useful for detecting the presence of a target object embedded within a noisy background, despite environmental perturbations and losses destroying the initial entanglement \cite{guha_gaussian-state_2009,pirandola_advances_2018,sanz_quantum_2017}. 
	
In 2013, Lopaeva et al. performed an experimental demonstration of the quantum illumination principle, to determine the presence or absence of a semi-transparent object, by exploiting intensity correlations of a quantum origin in the presence of thermal light \cite{lopaeva_experimental_2013}. Additionally, a quantum illumination protocol has been experimentally demonstrated in the microwave domain \cite{barzanjeh_microwave_2015} and a further demonstration in which joint detection of the signal and idler is not required \cite{chang_quantumenhanced_2019}. However, these previous demonstrations were restricted to simply detecting the presence or absence of a target, rather than performing any form of spatially resolved imaging.The acquisition of an image using quantum illumination has recently been reported \cite{england_quantumenhanced_2019}, but that demonstration was performed using a single-transverse-mode source of correlated photon pairs and by raster-scanning the object within one of these single-mode beams. The aforementioned demonstration may be seen as a qualitative assessment of the method but a full field imaging implementation of the quantum illumination protocol remains to be demonstrated using a multi-transverse-mode source of single photons and spatially resolved coincidence detections. Furthermore, the work in \cite{england_quantumenhanced_2019} uses only a coherent state as the as the incoming background light rather than a thermal state as was used in \cite{lopaeva_experimental_2013}. The use of thermal light as the incoming background light, as does the work presented here, better represents environmental light statistics and therefore performing full-field imaging using quantum illumination would be a demonstration of the potential real-world applications of the quantum illumination protocol.
	
In quantum imaging, commonly used properties are spatial quantum-correlations, which can be exploited to surpass the classical limits of imaging \cite{genovese_real_2016,lugiato_quantum_2002, walborn_spatial_2010,meda_photon-number_2017,Moreau2018review}. These quantum-correlations have been used in the case of NOON states for enhanced phase detection \cite{giovannetti_quantum-enhanced_2004,ono_entanglement-enhanced_2013}, through the use of definite number of photons, to improve the signal to noise ratio for measuring the absorption of objects through sub-shot-noise measurements \cite{meda_photon-number_2017,brida_experimental_2010,moreau_demonstrating_2017,toninelli_sub-shot-noise_2017}, and to perform resolution-enhanced imaging by centroid estimation of photon-pairs \cite{toninelli_resolution-enhanced_2019}. Such schemes rely on the ability to detect and utilise quantum proprieties after the probed object, and are therefore sensitive to decoherence through the introduction of environmental noise and optical losses that lead to severe degradation of the quantum enhancement \cite{giovannetti_advances_2011}. These schemes are therefore limited to low-noise and high-detection efficiency conditions, and to the sensing of objects presenting relatively low absorption. These limitations make such schemes difficult to implement in real-world conditions. However, the quantum illumination protocol, as described by Lloyd \cite{lloyd_enhanced_2008}, is resilient to environmental noise and losses. 
	
In this work we utilise spatial correlations between downconverted photon-pairs to demonstrate the first quantum illumination full-field imaging protocol. The spatial quantum correlations are manifested in the entangled photon-pairs produced via spontaneous parametric downconversion (SPDC). The twin beams produced via the SPDC process are directed to different regions of an Electron Multiplying CCD (EMCCD) array detector. EMCCD array detectors have previously been used to measure spatial correlations between photon-pairs \cite{moreau_realization_2012,edgar_imaging_2012,defienne_general_2018}. By performing a pixel-by-pixel AND-operation between two regions of an array detector containing the SPDC beams, we preferentially select correlated photon-pair events and reject uncorrelated background light and sensor noise events. As a result of this preferential selection, the quantum illumination AND-image resulting from the sum of these AND-events produces a contrast advantage relative to the classically acquired image, comprising the simple sum of the raw events. This relative contrast advantage and background rejection is achieved through the suppression of both background light and sensor noise. We also show that the advantage of the quantum illumination protocol over a classical imaging scheme is maintained in the presence of both noise and losses. 
The quantum advantage that we report is relative to the optimum classical measurement obtained via photon-counting and not obtained via any phase-sensitive measurement. This is in order to perform the demonstration using equivalent experimental conditions to acquire both the classical image and the quantum illumination AND-image in which an identical optical system, illumination source, illumination level, and noise, are used to collect data as it is acquired within the same acquisition. A phase-sensitive detection method \cite{tan_quantum_2008,guha_gaussian-state_2009} would require a different optical setup for the classical acquisition in addition to phase stability, and it would prove impractical to perform imaging in real world scenarios.
	
This demonstration of a quantum illumination full-field imaging protocol is an important development in the application of quantum technologies within a real-world setting in which background light and sensor noise requires suppression. 
Applications of quantum illumination protocols such as quantum LIDAR will become more viable as multi-pixel SPAD arrays that are capable of time-tagging single-photon events are developed. As opposed to the EMCCD camera we use here, an Andor iXon Ultra 888 with a single-frame acquisition time of $\sim 0.015$ s for our acquisition parameters (see the materials and methods section for full details regarding camera acquisition parameters), the current generation of SPAD arrays being developed have increased acquisition speed and time resolution with frame rates of 100 kfps and gate times of 4 ns respectively being achievable \cite{gyongy_256_201}. An improvement of around seven orders of magnitude in the temporal resolution of the detector will potentially allowing useful depth measurements of a scene to be made.
	
\section*{Results}
\textbf{Imaging system.} The experimental configuration is shown in Fig. 1. A 3 mm thick $\beta$-barium borate (BBO) non-linear crystal cut for type II degenerate downconversion is pumped by a collimated $\sim 8 \text{ mm}$ diameter laser beam at 355 nm to generate SPDC photon-pairs centred on the degenerate wavelength of 710 nm. A type-II phase-matched downconversion source results in the spatial separation of the two emitted SPDC beams labelled as the probe beam and the reference beam. The probe beam illuminates the object and may be subject to environmental losses and noise, while the reference beam neither interacts with the object nor is it subject to environmental losses and noise. After transmission through the crystal the pump beam is removed using a pair of dichroic mirrors and a pair of high-transmission interference filters mounted prior to the camera select the downconverted photon-pairs centred at the degenerate wavelength of 710nm. As shown in Fig. 1, the target object is placed in the far-field of the crystal such that the probe beam interacts with it while the reference beam has a free optical path. This plane is demagnified by a factor of eight and imaged onto an EMCCD array detector, such that the two downconverted beams are imaged by different regions of the EMCCD array detector of pixel size $13 \times 13 \text{ } \mu \text{m}^{2}$. The size of the correlations between the probe and reference beam in the far-field of the downconversion crystal depends upon the wavelength and diameter of the pump beam, and the effective focal length of the transform lens ($L_{1}$). For the configuration presented here this gives a correlation width of $\sigma \approx 4 \text{ } \mu \text{m}$ in the plane of the EMCCD camera chip. A background light field is deliberately introduced using a thermal light source to illuminate a mask which overlays the probe beam through a reflection from a microscope slide slip cover (MS) placed in the image plane of the crystal. A thermal light source is used so as to simulate real-world conditions in which environmental noise will follow a similar distribution (i.e. noisier than a Poisson distribution) and be broad-band illumination. The level of demagnification for the quantum imaging arm was set to maximise the number of SPDC photon-pairs arriving in anti-correlated detector pixel positions while maintaining the ability to construct an image which is not under sampled. See the materials and methods section for full details regarding optical setup and camera acquisition parameters.

\begin{figure}[H]
	\centering
	\includegraphics[width=14cm]{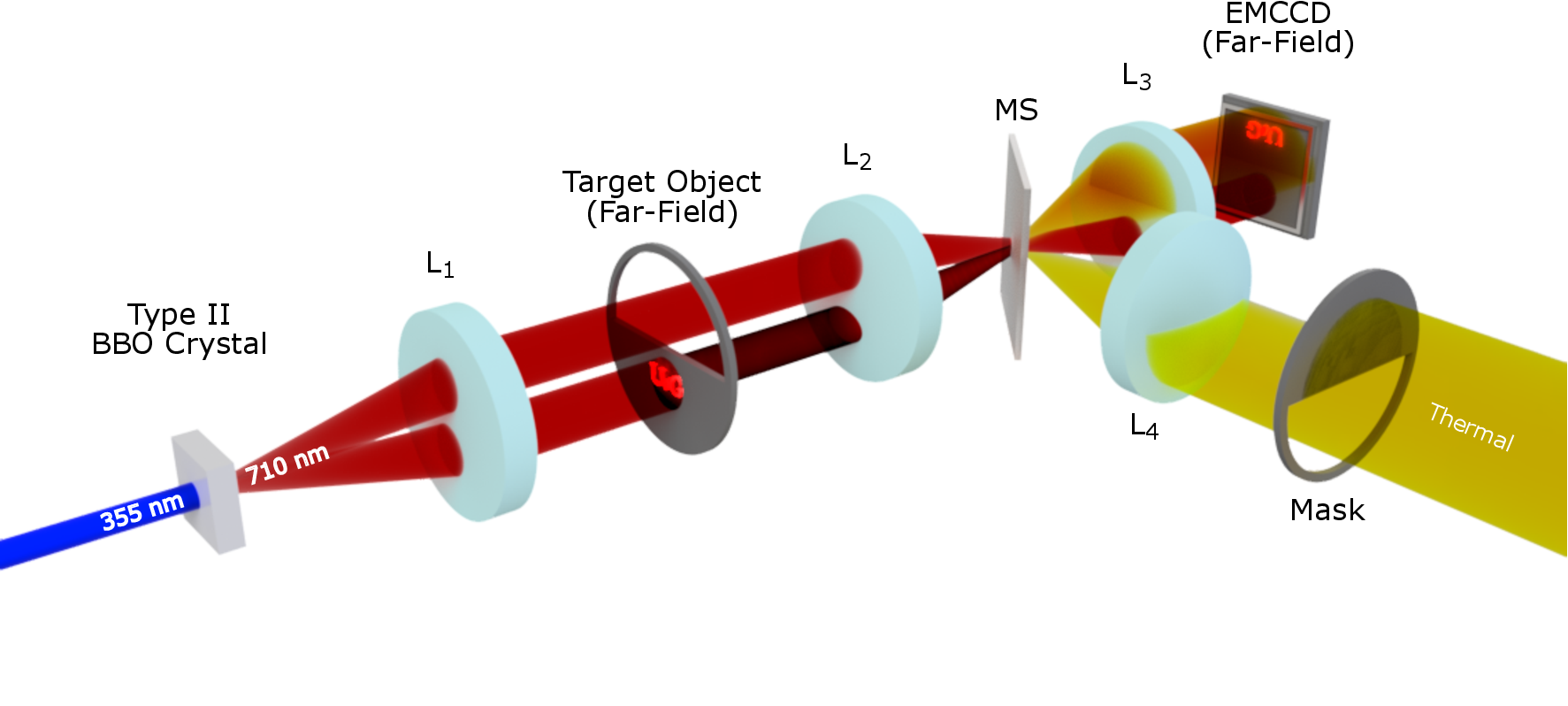}
	\caption{\textbf{Schematic of the quantum illumination experimental setup.} A BBO crystal cut for type-II downconversion is pumped by a UV laser to produce entangled photon-pairs via SPDC. The probe beam interacts with the Target Object placed in the far-field of the crystal while the reference beam has a free optical path. The lenses $ L_{1} = 75 \text{ mm, } L_{2} = 400\text{ mm, and } L_{3} = 50\text{ mm } $ comprise the quantum arm optics used to transform into the far-field, and then to demagnify by a factor of eight onto the EMCCD camera. A mask illuminated by thermal light is projected onto the camera using lenses $ L_{4} = 300 \text{ mm and } L_{3} = 50\text{ mm } $ by reflection from the microscope slide slip cover (MS) placed in the image plane of the crystal.}
	\label{fig:inksetup}
\end{figure}

We obtained our results using an illumination regime that was used by Tasca et al. (2013) \cite{tasca_optimizing_2013} in which the threshold events per pixel per frame from the detection of SPDC photons matches the event rate due to the noise of the camera which for our detector and acquisition settings is $ \sim0.0016 $ events per pixel per frame over the chosen acquisition region. As discussed above, background thermal illumination was added to the region of the sensor on which the probe beam is detected thereby simulating environmental noise at the probe beam wavelength which the quantum illumination protocol is able to reject. ND filters may be introduced after the target object to introduce optical losses as applied to the probe beam in order to demonstrate the resilience to losses of the quantum illumination protocol. 

\textbf{Image analysis.} To take advantage of the quantum illumination a simple pixel-by-pixel AND-operation between the two regions of the sensor that detect the reference and the probe beams within the same frame. This pixel-by-pixel AND-operation is equivalent to taking the Hadamard product of the two image arrays and is represented in Fig. 2 b) where it may be seen that those events that occur in diametrically opposite positions about the correlation peak are added together to build the quantum illumination AND-image. These events are selected by rotating the region of the camera frame corresponding to the probe beam through an angle of $\pi$ to transform the momentum anticorrelation into a position correlation and then performing the AND-operation to select the pixel coordinates in which an event is detected in both the reference and the probe beam. Performing this operation serves to select the momentum-anticorrelated photon-pairs which comprise the correlation peak presented in Fig. 2 a). The result of this software AND-operation over a number of frames is then summed to build the AND-image. Performing this pixel-by-pixel AND-operation on correlated events, as opposed to the classically acquired image that constitutes single events and also includes both sensor noise events and background illumination, an advantage in contrast is achieved. This advantage is due to the fact that the AND-operation will preferentially keep the photon-pair events and will reject most of the uncorrelated events that arise from either sensor noise or unwanted background illumination.

\begin{figure}[H]
	\centering
	\includegraphics[width=12cm]{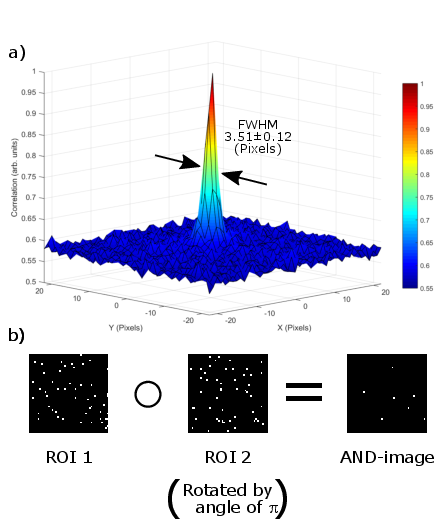}
	\caption{\textbf{Cross Correlation.} a) Correlation peak resulting from the cross-correlation of the two regions of interest that comprise the beams containing the photon-pairs. Calculated from 250,000 frames under an illumination regime identical to that in which the data was taken. b) Representation of the pixel-by-pixel AND-operation on a sum of 10 frames for visualisation purposes. Of the two regions of interest that comprise the regions of the frame on which the reference and probe beams are detected (ROIs) one must be rotated through an angle of $\pi$ in order to transform the momentum anticorrelation into a position correlation such that the AND-operation selects the momentum anticorrelated photon-pairs.  }
	\label{fig:MilesHat}
\end{figure}

The number of accidental coincidences between uncorrelated probe and reference events may be reduced by operating in a photon-sparse regime ($ \ll1 $ events per pixel per frame). In addition, operating in such a low-light regime may be desirable due to requirements of non-intrusive detection. In the photon-sparse regime under which the system operates ($ \ll1 $ events per pixel per frame after thresholding) the intensity correlations between classical beams are extremely weak \cite{lopaeva_experimental_2013} and therefore the comparison we use in this work is between the AND-image obtained using quantum illumination and the direct classical image. Both the AND-image and the classical image are acquired in the same acquisition using an identical optical setup. A general discussion on the optimal operating regimes for QI protocols can be found in \cite{tan_quantum_2008} and in \cite{lopaeva_experimental_2013,meda_photon-number_2017} in the context of photon-counting strategies.
	
We show that the advantage of the quantum illumination protocol is that it is resilient to thermal background noise and to losses introduced into the probe arm. A theoretical description of the protocol can be found in the supplementary material. To perform a fair analysis of the results a distinction must be made between the background light and sensor noise, and the noise within the images that is due to shot-noise. The quantum illumination AND-images contain fewer events resulting from background light and sensor noise as evidenced by the removal of the cage in Fig. 3 but exhibit greater shot-noise $ \sqrt{n}/n $, for which $ n $ is the number of events, when compared to the classically acquired image. The aim of our quantum illumination protocol is not to demonstrate an SNR improvement that would rely on sub-shot-noise statistics such as that demonstrated in Brida et al (2010) \cite{brida_experimental_2010}. Rather, the aim of the quantum illumination protocol presented here is the rejection of background light and sensor noise which is achieved by the preferential rejection of uncorrelated events under the assumption of an unknown, potentially structured background. As discussed in \cite{meda_photon-number_2017}, under the assumption that the background is known or can be independently estimated, the quantum illumination protocol will exhibit an advantage relative to the classically acquired image only when the number of accumulated photons per pixel is relatively low. In the case of an unknown background it is not possible to algorithmically subtract the background and as a result any subtraction applied may confuse or even deliberately mislead an interpretation of the true image.
	
\textbf{Rejection of structured thermal illumination.} The quantum illumination protocol works with arbitrary structured and a-priori unknown environmental background, which illustrates that the quantum illumination protocol works when a lack of knowledge of the background does not permit any ad-hoc background subtraction. We demonstrate that this quantum illumination protocol may be used to separate an object illuminated by the probe beam from a mask illuminated with thermal light. In Fig. 3 the bird and the fish are illuminated by the probe beam and the cage and net are illuminated by thermal light. By performing the same AND-analysis as previously described on the acquired frames to preferentially reject uncorrelated background light and sensor noise events the quantum-illuminated bird and fish may be distinguished from the thermally illuminated cage and net respectively. 
	
In order to assess the distinguishability of the object we must consider both the image contrast and the signal to noise. For a final image comprising the quantum-illuminated target object, of average value, $\left\langle O\right\rangle$, above the dark regions of the image, and the thermally illuminated structured background, of average value, $\left\langle S\right\rangle$, above the dark regions of the image, the rejection of the structured illumination in the form of the cage or net may be characterised as background rejection, $R_{Q/C} = \left\langle O\right\rangle / \left\langle S \right\rangle $. The rejection ratio, $RR$, is taken to be the ratio of the values obtained from the quantum illumination AND-image, $R_{Q}$, and the classically acquired image, $R_{C}$, as per Equation \ref{eqn:NRRRat}. 
	
\begin{equation}
RR = \frac{R_{Q}}{R_{C}}
\label{eqn:NRRRat}
\end{equation}
	
While the above rejection ratio metric assesses the rejection of the thermal background illumination it does not, however, take into account the shot-noise on the quantum-illuminated target object regions $(\sigma_{O})$. This noise $(\sigma_{O})$ will also affect the ability to distinguish the target object $(O)$ from the thermally illuminated structured background $(S)$. We therefore define a figure of merit, $D_{Q/C}$, to take into account the distinguishability of quantum-illuminated target object $(O)$ against both the thermally illuminated structured background $(S)$ and the noise on the object regions $(\sigma_{O})$ as per Equation \ref{eqn:CNRRat}.
	
\begin{equation}
D_{Q/C} = \frac{\left\langle O\right\rangle}{\left\langle S\right\rangle + \sigma_{O} }
\label{eqn:CNRRat}
\end{equation}
	
Note that in the limit where a very large number of frames are acquired, and in the assumption that the noise $\sigma_{O} $ is purely due to the shot noise, that is $\sigma_{O} = \sqrt{O}$, then the contribution of this noise in both $D_{C}$ and $D_{Q}$ vanishes. For the classically acquired image the limit in which this noise vanishes is under the acquisition of a smaller number of frames than for the quantum illumination AND-image due to a greater number of events being kept per frame. Under the limit of the acquisition of a very large number of frames the distinguishabilities become equivalent to the respective background rejection, $R_{C}$ and $R_{Q}$, as defined previously. The maximum attainable image distinguishability advantage is therefore given by the reduction ratio, $RR$, and is achieved when a large number of frames are acquired, such that both the classical and quantum images are smooth.
	
\begin{figure}[H]
	\centering
	\includegraphics[width=15cm]{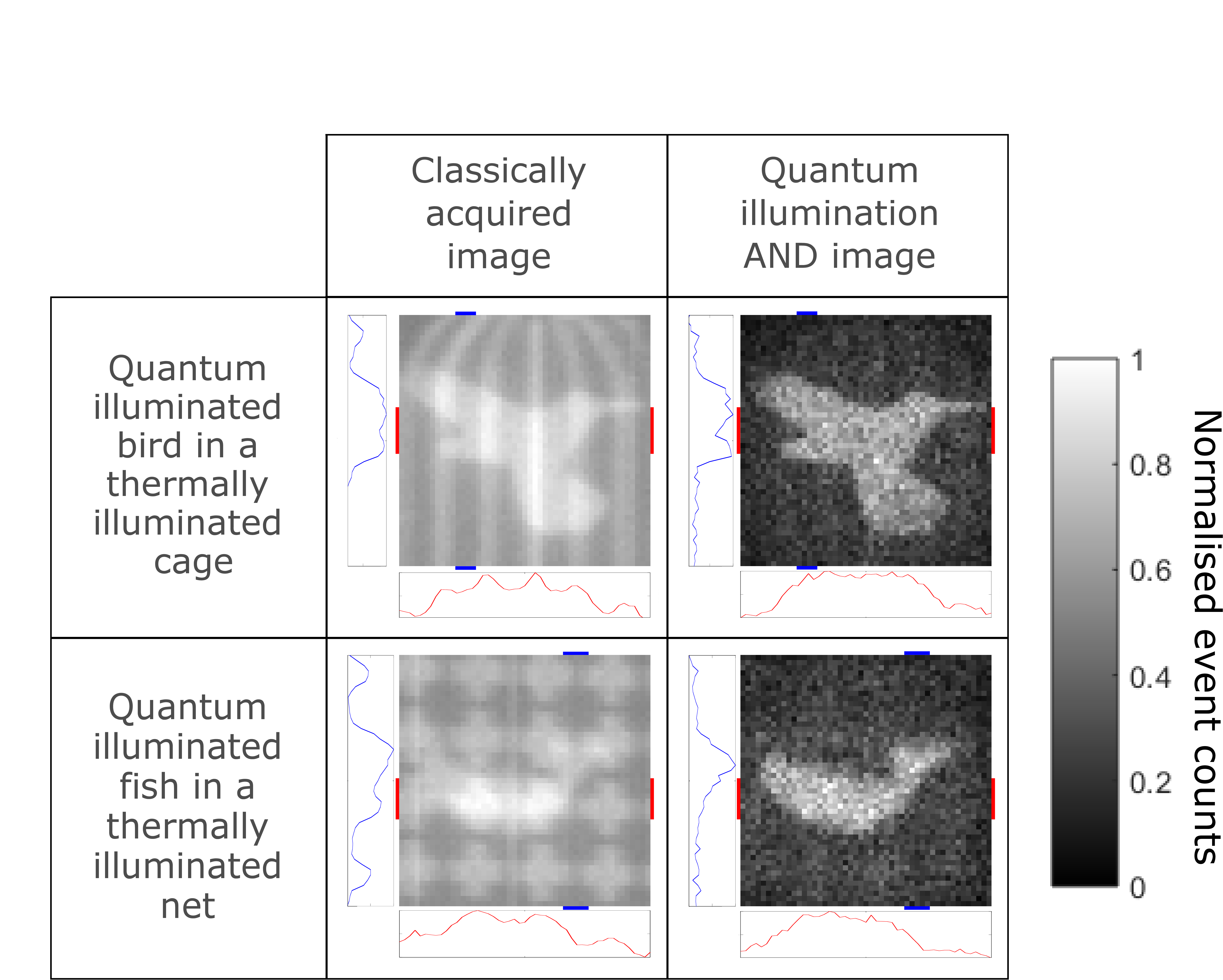}
	\caption{\textbf{Images of quantum illuminated target object preferentially selected over thermally illuminated mask.} Images of an object under quantum illumination overlaid with a thermally illuminated mask (second column). By applying the AND-operation on the data, the quantum illuminated object may be separated from the thermally illuminated mask (third column). In doing so the bird may be released from its cage and the fish released from its net. Red lines indicate the rows of the image used to generate the cut graph as shown below the images while blue lines indicate the columns used to generate the cut graph to the left of the images. Rows and columns used to generate the cut graphs as denoted by the red lines and blue lines respectively are rows 19-27 and columns 12-15 for the bird in a cage and rows 25-32 and columns 33-37 for the fish in a net. The scale of the cut graphs are normalised intensity (arbitrary units). Images are constructed over 2.5 million frames and are $49 \times 49$ pixels.}
	\label{fig:BirdCage}
\end{figure}

In Fig. 3 alongside the images horizontal and vertical cuts through the images are displayed in red and blue respectively. In the graphs referring to the classically acquired images there exist peaks corresponding to the thermally illuminated cage and net, while in the case of the graphs referring to the quantum illumination AND-images the prominence of these peaks are greatly reduced due to the preferential rejection of the uncorrelated background noise events. The graphs referring to the quantum illumination AND-image exhibit greater shot-noise due to the images containing fewer events than the classically acquired images and also some peaks remain identifiable in the cut graphs due to increased false correlations in local regions with an increased fill factor. For the bird in a cage as shown in Fig. 3 an improvement in the rejection ratio, RR, of a factor of order 4.2 is observed, and a value of 1.3 for the distinguishability metric $ D_{Q}/D_{C}$, mainly limited here by the efficiency of our implementation.
	
Figure 4 displays the bird in a cage under a range of increasing levels of thermal illumination. In all cases the prominence of the bars is reduced in the quantum illumination AND-image compared to the classically acquired image, however, some structure remains visible as a result of false correlations. The rejection of background and stray light for the images obtained under these illumination conditions as defined by the rejection ratio, $RR$, increases from 4.2 to 5.8 across as the level of thermal illumination increases. The distinguishability ratio $D_{Q} / D_{C}$ which takes into account the variations within the quantum illuminated object is also shown to present an advantage in the case of the quantum illumination AND-image when compared to the classically acquired image. The distinguishability value may also be seen to increase with increasing thermal illumination of the cage and this is due to the quantum illumination AND-image not keeping all of the classical cage events and therefore $D_{Q}$ will decrease at a slower rate than $D_{C}$ for the classically acquired image.

\begin{figure}[H]
	\centering
	\includegraphics[width=13cm]{figure4.eps}
	\caption{\textbf{Images of quantum illuminated target object preferentially selected over thermally illuminated mask with increasing thermal illumination and the introduction of losses.} Images of the bird under quantum illumination overlaid with a thermally illuminated cage (second column). By applying the AND-operation on the data, the quantum illuminated bird may be separated from the thermally illuminated cage (third column). In doing so the bird may be released from its cage. This is demonstrated across a range of increasing levels of thermal illumination represented in the first column as the ratio of the detected average intensity of the thermally illuminated cage image regions to the intensity of the quantum illuminated bird image regions. The final row presents results with the further introduction of $ \sim 50 \%$ optical losses into the probe beam by way of an ND filter. The distinguishability ratio $D_{Q} / D_{C}$ is presented in the fourth column and increases with increasing thermal illumination of the cage. Red lines indicate the rows of the image used to generate the cut graph as shown below the images while blue lines indicate the columns used to generate the cut graph to the left of the images. Rows and columns used to generate the cut graphs as denoted by the red lines and blue lines respectively are rows 19-27 and columns 12-15 for the images. The scale of the cut graphs are normalised intensity (arbitrary units). Images are constructed over 2.5 million frames and are $49 \times 49$ pixels.}
	\label{fig:BirdCageFull}
\end{figure}
	
\textbf{Quantum illumination advantage.} Having demonstrated the use of the QI protocol in terms of rejection of background we now go on to assess the robustness of this quantum illumination protocol under differing loss and thermal noise levels and measure the ability of our QI scheme to reject the background thermal light and sensor noise. Here we apply a unstructured background so as to simplify the demonstration compared to a structured background as used above. We use the ratio between the contrasts of the quantum illumination AND-image and the classical image. In both cases, image contrast is quantitatively assessed using the Michelson contrast or visibility ($V$) as represented in Equation (\ref{eqn:mc}), in which the difference between the calculated intensities of the bright region ($I_{max}$) and the dark region ($I_{min}$) of the final summed image is divided by the sum of the intensities in the bright and dark regions.
	
\begin{equation}
V_{Q/C} = \frac{ I_{max} - I_{min} }{ I_{max} + I_{min} }
\label{eqn:mc}
\end{equation} 
	
To assess the advantage of the quantum illumination protocol, in this work we define the quantum illumination advantage, $A$, as the ratio between the contrast $V_Q$ of the image acquired through quantum illumination (i.e. the image obtained by keeping only the correlated events through performing the AND-operation) and the contrast $V_C$ of the image acquired simply by summing the probe-beam events over all frames.
	
\begin{equation}
A = \frac{ V_{Q} }{ V_{C} }
\label{eqn:contrastadv}
\end{equation} 
	
So as to verify that the scheme presented here is a quantum illumination scheme a binary mask of the University of Glasgow (UoG) initials laser cut from a piece of card was used as the target object. Figure 5 shows a comparison between performing the aforementioned AND-analysis on 1.5 million frames to the corresponding classical image over a range of increasing optical losses introduced into the probe arm of the system by means of a series of ND filters.

\begin{figure}[H]
	\centering
	\includegraphics[width=15cm]{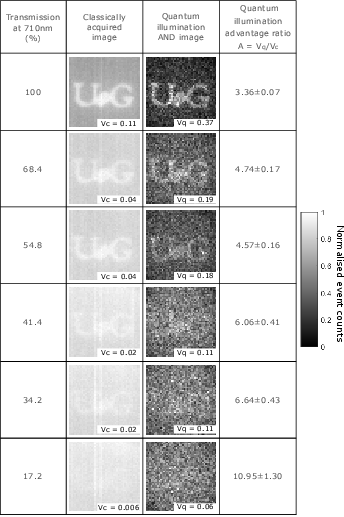}
	\caption{\textbf{Imaging using quantum illumination with losses introduced.} Imaging using quantum illumination within a thermal background with optical losses introduced. Images of the UoG object with the classical image created by averaging all frames (second column), and the quantum illumination AND-image built from the sum of results of performing an AND-operation to select correlated events in the reference and probe beams (third column). The UoG object illuminated by the probe beam is imaged under the conditions of increasing optical losses (see the transmission of the ND filters placed into the probe beam at 710nm in the first column). The quantum illumination advantage A under these levels of optical losses for images constructed over 1.5 million frames is displayed in the fourth column. Images are $45 \times 45$ pixels.}
	\label{fig:UoGLosses}
\end{figure}
	
From the images presented in Fig. 5 it is seen that the quantum illumination advantage, $A$, increases with the level of optical losses introduced. This increase in the quantum illumination advantage is due to the fact that the classical image contrast, in which all events are kept, degrades faster than that of the quantum illumination AND-image constructed using the result of the AND-operation (see supplementary material for a theoretical description of the effect).
The quantum illumination advantage also increases with the level of background noise events introduced into the probe beam from a thermal source, as is the expected behaviour from the equations we present in the supplementary material. We also present the corresponding experimental results obtained under increasing levels of thermal illumination introduced into the system in the supplementary material. 
	
In addition to the imaging results for this quantum illumination protocol we also consider how it may be used in an application where the presence or absence of an object needs to be assessed. This is the context in which Lloyd \cite{lloyd_enhanced_2008} originally proposed the quantum illumination protocol and has clear applications in realising quantum LIDAR or quantum RADAR applications. It can be demonstrated that a strategy combining the information from the classical data and the quantum data, both acquired through quantum illumination, will always lead to a bit-error-rate enhancement, as long as a contrast advantage is detected in the quantum data. This bit-error-rate enhancement in the context of quantum illumination protocols is discussed in the supplementary material.
	
\section*{Discussion}
We have demonstrated a quantum illumination protocol to perform full-field imaging achieving a contrast enhancement through the suppression of both background light and sensor noise. Structure within the thermal background illumination is potentially a-priori unknown and therefore cannot be suppressed with a simple ad-hoc background subtraction. Through resilience to environmental noise and losses, such a quantum illumination protocol should find applications in real-world implementations including quantum microscopy for low light-level imaging, quantum LIDAR imaging applications, and quantum radar. Improvements in detector technologies such as SPAD arrays capable of time-tagging events should enable time-of-flight applications to be realised and applied outside of the laboratory through the increased acquisition speed and time resolution that they enable.

\section*{Materials and Methods}
The source used for the pump is a JDSU xCyte CY-355-150 Nd:YAG laser with $355 \text{ nm}$ output at 160 mW with pulse repetition 100 +/- 10 Mhz. The pump is attenuated to $\sim 1 \text{ mW mm}^{-2}$ and expanded to $ \sim 8 \text{ mm}$ on the crystal via a spatial filter comprising a $50 \text{ mm}$ lens, $25 \mu\text{m}$ pinhole, and a $200 \text{ mm}$ lens. The downconversion source is a $\beta$-barium borate (BBO) non-linear crystal of dimensions $10 \times 10\times 3 \text{ mm}$, cut for type-II degenerate downconversion at $710 \text{ nm}$ with a half-opening angle of $5^{\circ}$. The camera used here is an Andor iXon ULTRA 888 DU-888U3-CS0-\#BV; of pixel size $13\times13 \mu\text{ m}^2$, $100\%$ fill-factor EMCCD camera. The camera was cooled to $-90^{\circ}\text{C}$ using liquid cooling. Optimal acquisition parameters for the camera were set as follows: vertical speed $1.33 \mu\text{s}$; voltage clock amplitude $ +0 \text{ V}$, horizontal speed $10 \text{ MHz}$; EM gain $1000$; pre-amplifier gain set to $1$; $128 \times 128$ pixel acquisition region; exposure time of $0.0149500 \text{ s}$ (the shortest exposure time for given acquisition parameters).
	
The filters used in this experiment are Chroma T455lpxt dichroic mirrors ($455 \text{ nm}$ cut-off, $98\%$ transmission at $710 \text{ nm}$) and Chroma ET710/10m interference filters (centred on the degenerate wavelength $710 \text{nm}$ with a bandwidth of $10 \text{ nm}$ and a top-hat profile, $99\%$ transmission at $710 \text{ nm}$). Two interference filters are used as degenerate downconversion is not centered on the bandpass thereby resulting in unevenly sized downconversion beams. Further to placing one filter onto the camera the a second interference filter is orientated to shift its transmission profile and so select a narrower overall bandpass and more evenly sized beams.
	
To each of the frames acquired by the EMCCD camera a threshold is applied so as to minimise the camera readout noise events which dominates the low values of the Analog-Digital Converter (ADC) counts histogram for EMCCD cameras, while maximising the overall quantum efficiency of the system. A threshold of $T \approx \mu_{ro} + 3 \text{ } \sigma_{ro}$ is appropriate in which $\mu_{ro}$ and $\sigma_{ro}$  are the mean and standard deviation of the readout noise respectively. Any frames containing events of an extraordinary nature, taken to be those with pixel values of greater than 45,000, were removed from the dataset due their containing events caused by cosmic rays striking the detector. See also \cite{toninelli_sub-shot-noise_2017}. For the acquisition parameters used here a threshold of $510$ is applied and this results in $\sim 0.0016$ camera noise events per pixel per frame.


\begin{thebibliography}{10}
	
	\bibitem{lloyd_enhanced_2008}
	S.~Lloyd, Enhanced sensitivity of photodetection via quantum illumination. {\it
		Science\/} {\bf 321}, 1463--1465 (2008).
	
	\bibitem{tan_quantum_2008}
	S.-H. Tan, B.~I. Erkmen, V.~Giovannetti, S.~Guha, S.~Lloyd, L.~Maccone,
	S.~Pirandola, J.~H. Shapiro, Quantum {Illumination} with {Gaussian} {States}.
	{\it Physical Review Letters\/} {\bf 101}, 253601 (2008).
	
	\bibitem{shapiro_defeating_2009}
	J.~H. Shapiro, Defeating passive eavesdropping with quantum illumination. {\it
		Physical Review A\/} {\bf 80}, 022320 (2009).
	
	\bibitem{zhang_entanglements_2013}
	Z.~Zhang, M.~Tengner, T.~Zhong, F.~N.~C. Wong, J.~H. Shapiro, Entanglement’s
	{Benefit} {Survives} an {Entanglement}-{Breaking} {Channel}. {\it Physical
		Review Letters\/} {\bf 111}, 010501 (2013).
	
	\bibitem{guha_gaussian-state_2009}
	S.~Guha, B.~I. Erkmen, Gaussian-state quantum-illumination receivers for target
	detection. {\it Physical Review A\/} {\bf 80}, 052310 (2009).
	
	\bibitem{pirandola_advances_2018}
	S.~Pirandola, B.~R. Bardhan, T.~Gehring, C.~Weedbrook, S.~Lloyd, Advances in
	photonic quantum sensing. {\it Nature Photonics\/} {\bf 12}, 724--733 (2018).
	
	\bibitem{sanz_quantum_2017}
	M.~Sanz, U.~Las~Heras, J.~García-Ripoll, E.~Solano, R.~Di~Candia, Quantum
	{Estimation} {Methods} for {Quantum} {Illumination}. {\it Physical Review
		Letters\/} {\bf 118}, 070803 (2017).
	
	\bibitem{lopaeva_experimental_2013}
	E.~D. Lopaeva, I.~R. Berchera, I.~P. Degiovanni, S.~Olivares, G.~Brida,
	M.~Genovese, Experimental realisation of quantum illumination. {\it Physical
		Review Letters\/} {\bf 110}, 153603 (2013).
	
	\bibitem{barzanjeh_microwave_2015}
	S.~Barzanjeh, S.~Guha, C.~Weedbrook, D.~Vitali, J.~H. Shapiro, S.~Pirandola,
	Microwave quantum illumination. {\it Physical Review Letters\/} {\bf 114},
	080503 (2015).
	
	\bibitem{chang_quantumenhanced_2019}
	C.~W. Sandbo Chang, A.~M. Vaddiraj, J.~ Bourassa, B.~ Balaji, C.~M. Wilson, Quantum-enhanced noise radar. {\it Applied Physics Letters\/} {\bf 114}, 112601 (2019).
	
	\bibitem{england_quantumenhanced_2019}
	D.~G. England, B.~ Balaji, B.~J. Sussman, Quantum-enhanced standoff detection using correlated photon pairs. {\it Physical Review A\/} {\bf 99}, 023828 (2019).
	
	\bibitem{genovese_real_2016}
	M.~Genovese, Real applications of quantum imaging. {\it Journal of Optics\/}
	{\bf 18}, 073002 (2016).
	
	\bibitem{lugiato_quantum_2002}
	L.~A. Lugiato, A.~Gatti, E.~Brambilla, Quantum imaging. {\it Journal of Optics
		B: Quantum and Semiclassical Optics\/} {\bf 4}, S176 (2002).
	
	\bibitem{walborn_spatial_2010}
	S.~P. Walborn, C.~H. Monken, S.~Pádua, P.~H. Souto~Ribeiro, Spatial
	correlations in parametric down-conversion. {\it Physics Reports\/} {\bf
		495}, 87--139 (2010).
	
	\bibitem{meda_photon-number_2017}
	A.~Meda, E.~Losero, N.~Samantaray, F.~Scafirimuto, S.~Pradyumna, A.~Avella,
	I.~Ruo-Berchera, M.~Genovese, Photon-number correlation for quantum enhanced
	imaging and sensing. {\it Journal of Optics\/} {\bf 19}, 094002 (2017).
	
	\bibitem{Moreau2018review}
	P.-A. Moreau, E.~Toninelli, T.~Gregory, M.~J. Padgett, Imaging with quantum
	states of light. {\it Nature Reviews Physics\/} {\bf 1}, 367-380 (2019).
	
	\bibitem{giovannetti_quantum-enhanced_2004}
	V.~Giovannetti, S.~Lloyd, L.~Maccone, Quantum-{Enhanced} {Measurements}:
	{Beating} the {Standard} {Quantum} {Limit}. {\it Science\/} {\bf 306},
	1330--1336 (2004).
	
	\bibitem{ono_entanglement-enhanced_2013}
	T.~Ono, R.~Okamoto, S.~Takeuchi, An entanglement-enhanced microscope. {\it
		Nature Communications\/} {\bf 4}, 2426 (2013).
	
	\bibitem{brida_experimental_2010}
	G.~Brida, M.~Genovese, I.~R. Berchera, Experimental realization of
	sub-shot-noise quantum imaging. {\it Nature Photonics\/} {\bf 4}, 227--230
	(2010).
	
	\bibitem{moreau_demonstrating_2017}
	P.-A. Moreau, J.~Sabines-Chesterking, R.~Whittaker, S.~K. Joshi, P.~M.
	Birchall, A.~McMillan, J.~G. Rarity, J.~C.~F. Matthews, Demonstrating an
	absolute quantum advantage in direct absorption measurement. {\it Scientific
		Reports\/} {\bf 7}, 6256 (2017).
	
	\bibitem{toninelli_sub-shot-noise_2017}
	E.~Toninelli, M.~P. Edgar, P.-A. Moreau, G.~M. Gibson, G.~D. Hammond, M.~J.
	Padgett, Sub-shot-noise shadow sensing with quantum correlations. {\it Optics
		Express\/} {\bf 25}, 21826 (2017).
	
	\bibitem{toninelli_resolution-enhanced_2019}
	E.~Toninelli, P.-A.~Moreau, T.~Gregory, A.~Mihalyi, M.~P. Edgar, N.~Radwell, M.~J.
	Padgett, Resolution-enhanced quantum imaging by centroid estimation of biphotons. {\it Optica\/} {\bf 6}, 347-353 (2019).
	
	\bibitem{giovannetti_advances_2011}
	V.~Giovannetti, S.~Lloyd, L.~Maccone, Advances in quantum metrology. {\it
		Nature Photonics\/} {\bf 5}, 222--229 (2011).
	
	\bibitem{moreau_realization_2012}
	P.-A.~Moreau, J.~Mougin-Sisini, F.~Devaux, E.~Lantz, Realisation of the purely spatial Einstein-Podolsky-Rosen paradox in full-field images of spontaneous parametric down-conversion. {\it Physical Review A\/} {\bf 86}, 010101 (2012).
	
	\bibitem{edgar_imaging_2012}
	M.~P. Edgar, D.~S. Tasca, F.~Izdebski, R.~E. Warburton, J.~Leach, M.~Agnew, G.~S. Buller, R.~W. Boyd, M.~J. Padgett, Imaging high-dimensional spatial entanglement with a camera. {\it Nature Communications\/} {\bf 3}, 984 (2012).
	
	\bibitem{defienne_general_2018}
	H.~Defienne, M.~Reichert, J.~W. Fleischer, General model of photon-pair detection with an image sensor. {\it Physical Review Letters\/} {\bf 120}, 203604 (2018).
	
	\bibitem{gyongy_256_201}
	I.~Gyongy, N.~Calder, A.~Davies, N.~A.W. Dutton, R.~R. Duncan, C.~Rickman, P.~Dalgarno, R.~K. Henderson, A $256 \times 256$, 100-kfps, 61\% fill-Factor {SPAD} image sensor for time-resolved microscopy applications. {\it IEEE Transactions on Electron Devices\/} {\bf 65}, 547-554 (2018).
	
	\bibitem{tasca_optimizing_2013}
	D.~S. Tasca, M.~P. Edgar, F.~Izdebski, G.~S. Buller, M.~J. Padgett, Optimizing
	the use of detector arrays for measuring intensity correlations of photon
	pairs. {\it Physical Review A\/} {\bf 88}, 013816 (2013).
	
	
	
\end{thebibliography}

\noindent \textbf{Acknowledgements:} \\ 
\textbf{Funding:} This work was funded by the UK EPSRC (QuantIC EP/M01326X/1) and the ERC (TWISTS, Grant no.340507). T.G. acknowledges the financial support from the EPSRC (EP/N509668/1) and the Professor Jim Gatheral Quantum Technology Studentship. P.A.M. acknowledges the support from the European Union Horizon 2020 research and innovation program under the Marie Sklodowska-Curie fellowship grant agreement No 706410, from the Leverhulme Trust through the Research Project Grant ECF-2018-634 and from the Lord Kelvin Adam Smith Leadership Fellowship Scheme. E.T. acknowledges the financial support from the EPSRC Centre for Doctoral Training in Intelligent Sensing and Measurement (EP/L016753/1).

\includepdf[pages=-]{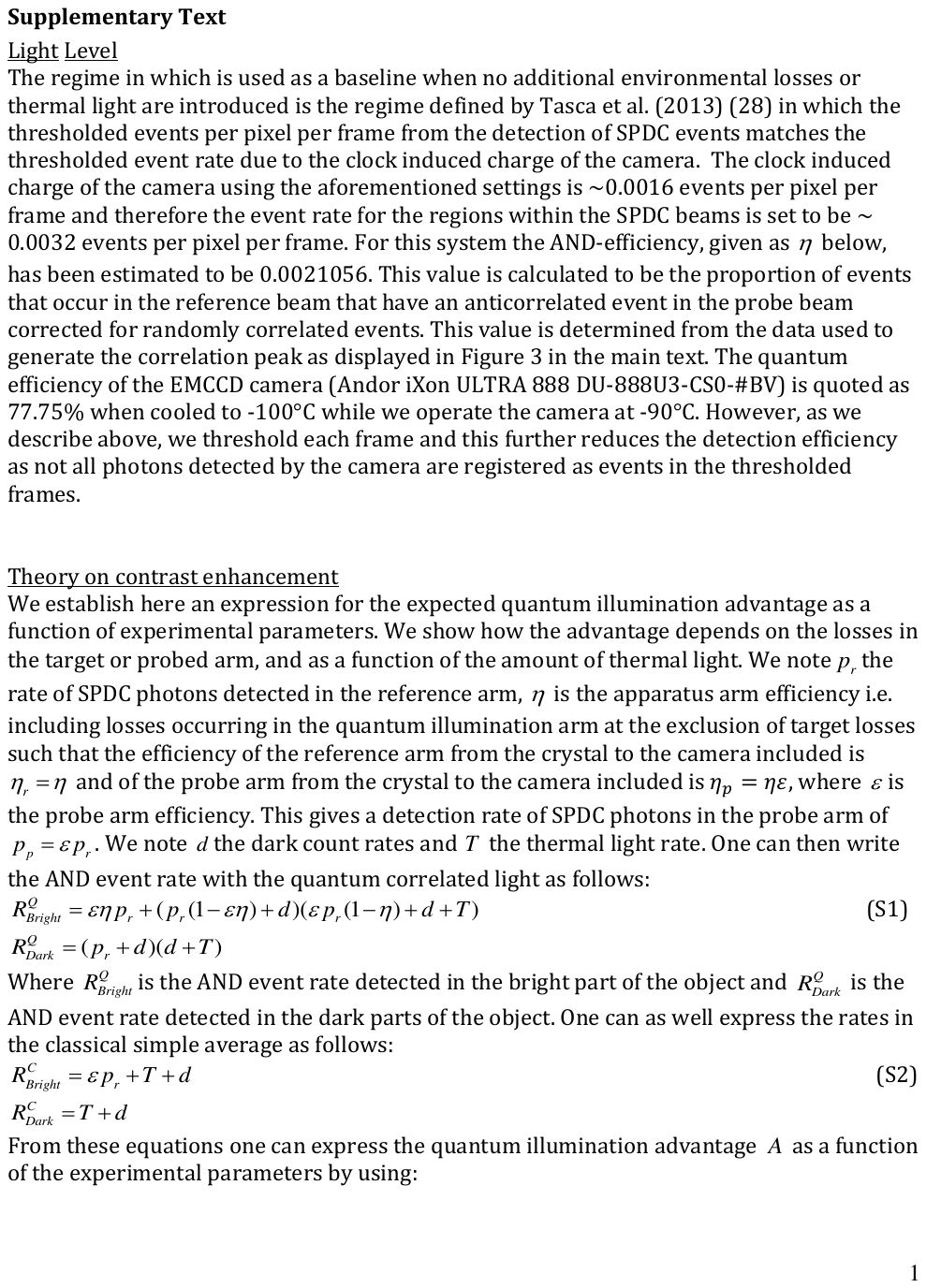}

\end{document}